\documentclass[useAMS,usegraphicx,usenatbib]{mn2e}
\usepackage{journals}
\usepackage{color}
\usepackage{longtable}
\usepackage[table]{xcolor}
\usepackage{supertabular}
\usepackage{colortbl}
\usepackage{rotating}
\usepackage{lscape}
\usepackage{times}
\usepackage[utf8]{inputenc}
\usepackage{mathtools}
\usepackage{ulem}

\begin{document}
\definecolor{darkred}{rgb}{0.5,0.0,0.0}
\definecolor{darkgreen}{rgb}{0.0,0.5,0.0}
\newcommand{\euge}[1]{{\cellcolor{gray!50}\textcolor{darkgreen}{Euge: #1}}}
\newcommand{\ariel}[1]{{\cellcolor{gray!50}\textcolor{darkred}{Ariel: #1}}}
\newcommand{\ct}[1]{{\cellcolor{gray!50}#1}}
\newcommand{\plh}{%
  {\ooalign{$\phantom{0}$\cr\hidewidth$\scriptstyle\times$\cr}}%
}
\newcommand*{\x}{\mathsf{x}\mskip1mu}
\title [CGs in different photometric bands]
{On the properties of compact groups identified in different photometric bands}
\author[Taverna et al.]
{Antonela Taverna$^{1}$\thanks{antotaverna@gmail.com},
Eugenia D\'iaz-Gim\'enez$^{1,2}$, 
Ariel Zandivarez$^{1,2}$,
\and
Francisco Joray$^{3}$, 
Mar\'ia Jos\'e Kanagusuku$^{1}$
\\
\\
$1$ Instituto de Astronom\'ia Te\'orica y Experimental (IATE), UNC, CONICET, OAC, Córdoba, Argentina. \\
$2$ Observatorio Astron\'omico (OAC), Universidad Nacional de C\'ordoba (UNC), Laprida 854, X5000BGR, 
C\'ordoba, Argentina.\\
$3$ Facultad de Matem\'atica, Astronom\'ia y F\'isica (FaMAF), UNC.\\
}

\date{\today}
\pagerange{\pageref{firstpage}--\pageref{lastpage}}
\maketitle
\label{firstpage}

\begin{abstract}
Historically, compact group catalogues vary not only in their identification algorithms and selection functions, but also in their photometric bands. Differences between compact group catalogues have been reported. However, it is difficult to assess the impact of the photometric band in these differences given the variety of identification algorithms. We used the mock lightcone built by \cite{Henriques_2012} to identify and compare compact groups in three different photometric bands: $K$, $r$, and $u$. We applied the same selection functions in the three bands, and found that compact groups in the u-band look the smallest in projection, the difference between the two brightest galaxies is the largest in the K-band, while compact groups in the r-band present the lowest compactness. We also investigated the differences between samples when galaxies are selected only in one particular band ({\it pure} compact groups) and those that exist regardless the band in which galaxies were observed ({\it common} compact groups). We found that the differences between the total samples are magnified, but also some others arise: pure-r compact groups are the largest in projection; pure-u compact groups have the brightest first ranked galaxies, and the most similar two first ranked galaxies; pure-K compact groups have the highest compactness and the most different two first ranked galaxies; and common compact groups show the largest percentage of physically dense groups. Therefore, without a careful selection and identification of the samples, the characteristic features of group properties in a particular photometric band could be overshadowed.
\end{abstract}


\section{Introduction}
Given their extreme nature, compact groups (CGs) are one the favourite
objects in extragalactic astronomy to study galaxy formation and evolution. 
Galaxy interactions are supposed to occur more likely within 
these small system of galaxies, 
hence important clues about galaxy evolution can be obtained
from the analysis of the physical properties of the galaxy members and 
their host groups.

During the last fifty years, several attempts have been done to construct
CGs catalogues, providing the possibility of performing comparisons of the CG
physical properties among different catalogues. 
Starting with the pioneer attempt of \cite{rose77}
and the well known catalogue of \cite{hickson82}, several other
authors have embarked themselves in the work of building CGs
catalogues, such as 
\cite{prandoni94,mcconnachie09,Diaz_Gimenez+12} and 
\cite{herfer15}. We mention only these particular catalogues because they 
are examples of surveys characterised by galaxy detections with different photometry:
B \citep{ubv}, R \citep{vri}, $b_j$ (COSMOS-UKST, \citealt{ukst}), 
$r$ (Sloan Digital Sky Survey, SDSS, \citealt{sdss}), $K_s$ (Two Micron All-Sky Survey, 
2MASS, \citealt{2mass}) and FUV (Galaxy Evolution Explorer, GALEX, \citealt{galex}) bands.
All these surveys span a wide range of the electromagnetic spectrum 
(roughly from 1400 to 23000 \AA), implying that objects with very different physical 
properties could be detected depending on which part of the spectrum is adopted to 
construct the galaxy surveys. Also, the different selection functions of the surveys 
(e.g., apparent magnitude limits, sky coverage) 
and algorithms to identify CGs (e.g., visual or automatic search, 
Hickson-like or friends-of-friends - FoF - type, with or without velocity filter) 
are very uneven among the existing CG catalogues. Hence, all these issues contribute to make the 
comparison among different samples of CGs a difficult task. 

For instance, \cite{Diaz_Gimenez+12} compared the samples of CGs identified by themselves in the K-band
(automatic Hickson-like plus flux limit algorithm) with the samples of CGs identified in the B band 
(\citealt{focardi02}, FoF-like algorithm), and in the R-band (\citealt{allam00}, FoF algorithm; 
and \citealt{hickson92}, visual inspection). They found that CGs in the R-band have smaller projected sizes, 
projected intergalaxy separations and crossing times than the other two catalogues. 
They also found that the K-sample was the first to show
statistically large first–second ranked galaxy magnitude gap.  
\cite{herfer15} have also performed a comparative analysis between the CGs identified by themselves 
from ultraviolet sources (SFCG, FoF algorithm on the plane of sky) with the samples of CGs in the 
K-band \citep{Diaz_Gimenez+12}, R-band \citep{hickson92} and in 
the r-band (\citealt{mcconnachie09}, automatic Hickson-like algorithm). 
They found that the SFCGs present the lowest velocity dispersions (and virial masses), 
while the R-CGs present the smallest projected intergalaxy separations and crossing times. 
Nevertheless, given the differences in the identification processes, 
it is hard to tell which is the nature of those differences.
Therefore, a fair comparison among different samples of CGs is required.

Semi-analytic models of galaxy formation combined with N-body numerical simulations have proved 
to be efficient at reproducing the properties of CGs in different bands
(e.g., \citealt{mcconnachie09,Diaz_Gimenez+10}). Therefore, 
in this work we use an all-sky lightcone \citep{Henriques_2012}
constructed using the semi-analytic galaxies extracted from the Millennium Simulation
\citep{springel05} to identify CGs in three different photometric 
bands: $u$ (SDSS) , $r$ (SDSS) and $K_s$ (2MASS).  
This will allow us to standardise the conditions for the identification of CGs
and performing a comparison among the resulting samples of CGs. 
This should shed some 
light on whether those differences observed in previous observational 
identifications are caused by using different algorithms and restrictions to identify CGs,
an unfair or biased comparison among samples, or if CGs in different photometric bands are
intrinsically different. 

The layout of this work is as follows: we describe the mock catalogue,
the compact group identifications and the analysis of their properties
in Section 2. In Sect.~3, we split the sample of CGs into those that 
exist regardless the photometric band, 
and those that only exist in one particular band, 
and we compare their properties. Finally, in Section 4 we summarise our results and conclusions.

\section{The compact group samples}
\subsection{The mock catalogue}
We identified compact groups of galaxies in the publicly available 
all-sky lightcone built by \cite{Henriques_2012}\footnote{
Galaxy mock lightcone available as table wmap1.BC03\_AllSky\_001 
at http://www.mpa-garching.mpg.de/millennium/}. 
Those authors constructed the lightcone 
using the semi-analytic galaxies extracted from the \cite{Guo_2011} 
by replicating the Millennium simulation box ($500 \rm \ Mpc \ h^{-1}$ on a side), 
and selecting galaxies from the different outputs of the simulation 
to include the evolution of structures and galaxy properties with time.
The semi-analytic model of galaxy formation of \cite{Guo_2011} introduced several
modifications with respect to earlier models. 
For instance, they introduced changes in 
the treatments of gas accretion, SN feedback, sizes of galaxies, and stripping of gas and stars.
The implementation of these modifications led them to a model that reproduces 
very well the observed abundance and the large-scale clustering of low redshift 
galaxies as a function of stellar mass, luminosity and colour, and also reproduces the 
colour distribution and the small-scale clustering of SDSS galaxies.

\cite{Henriques_2012}'s original mock lightcone is limited to an apparent observer-frame 
AB magnitude of $i< 21.0$ and includes apparent observer-frame 
magnitudes for 9 filters: SDSS u, g, r, i, z and VISTA Y, J, H, Ks.
In this work, we focused in three of these bands: 
Ks\footnote{Hereafter, we will refer to this magnitude just as K instead of Ks}, r, and u. 

\begin{figure}
\begin{center}
\includegraphics[width=\hsize]{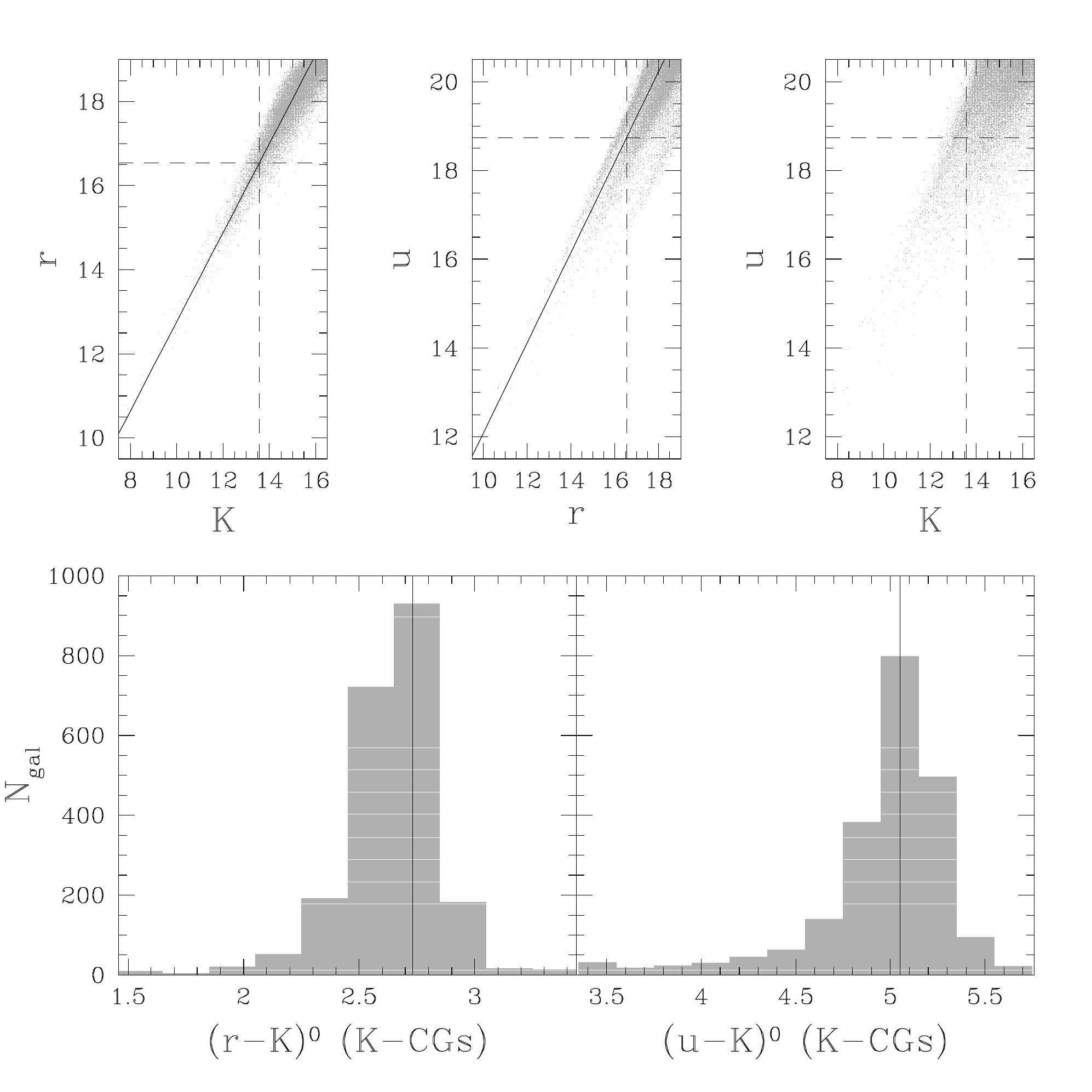}
\caption{\label{f1} Top panels: galaxy magnitudes in the parent lightcone 
(only 0.05\% of the original points are plotted). Solid lines represent the linear 
fit to the data, while dashed lines represent the limit adopted in each band.  
Bottom panels: distribution of differences between the rest frame apparent magnitudes in two different 
photometric bands of galaxies in Compact Groups identified in the K-band (Sect.~\ref{iden}). 
Solid lines represent the values adopted to determine the band shift to compute the surface brightness limit. 
}
\end{center}
\end{figure}
\begin{table}
\centering
\caption{Main parameters used to select galaxies in the mock catalogues, 
identify compact groups in different photometric bands, 
and assign sizes to the simulation particles.}
\begin{tabular}{cccc}
\hline
 & K & r & u \\
\hline
$m_{\rm lim}$ & 13.57  & 16.54 &  18.73\\
\# galaxies & 701449 & 677351 & 928367 \\
$<z>$ & 0.065 & 0.057 & 0.063 \\
\hline
$ m_{{\rm bri}_{\rm lim}} $ & 10.57 & 13.54 & 15.73  \\
Band shift & \nodata & r=K+2.73 & u=K+5.05 \\
$\mu_{\rm lim}$ &  23.6 & 26.33 & 28.65  \\
$M_\odot^{(1)}$ & 3.29  & 4.65 & 6.44 \\
\hline
$\alpha^{(2)}$     & $0.11$ & $0.11$ & $0.12$  \\
$\beta^{(2)}$      & $0.72$ & $0.78$ & $0.69$  \\
$\gamma^{(2)}$     & $0.09$ & $0.12$ & $0.11$  \\
${\cal M}_0^{(2)}$ & $1.57 \times 10^{10}$ & $2.25 \times 10^{10}$ & $1.70 \times 10^{10}$  \\
$a^{(2)} $ & $20.45 \times 10^{-3}$ & $37.24 \times 10^{-3}$ & $23.75 \times 10^{-3}$ \\
$b^{(2)} $ & $0.22$ & $0.20$ & $0.23$ \\

\hline
\end{tabular}
\label{tab1}
\parbox{8cm}{
\small
{\bf Notes.}
(1) www.ucolick.org/$\sim$cnaw/sun.html; 
(2) Prescriptions from \cite{Lange_2015} 
to compute the galaxy half-light radii ($R_i$) as a 
function of the stellar mass (${\cal M}^*_i$). 
Elliptical galaxies:
$R_i= \gamma \, ({\cal M}^*_i) ^\alpha(1+{\cal M}^*_i/{\cal M}_0)^{(\beta-\alpha)}$.
Non-elliptical galaxies: $R_i=a ({\cal M}^*_i)^b$ 

}
\end{table}

To mimic the observational sample of compact groups identified 
in a previous work in the 2MASS catalogue \citep{Diaz_Gimenez+12}, 
we converted the apparent K(AB) band available for the mock galaxies 
from the AB system to the Vega system to match the 2MASS magnitudes: 
$K({\rm Vega}) = K({\rm AB}) - 1.85$ \citep{cohen03,targett12}, 
and selected galaxies down to an apparent magnitude limit of $K = 13.57$. 
The sample comprises $701449$ galaxies within a solid angle of $4 \, \pi$. 
The number density of this particular mock galaxy catalogue reproduces the 
number density observed in the 2MASS catalogue remarkably well.
We also selected mock galaxy catalogues in the u- and in the r-bands. 
The apparent magnitude limits imposed in each of these bands were 
determined in order to mimic the limit in the K-band. 
Therefore, we examined the distribution of magnitudes in the three bands. 
Top panels in Fig.~\ref{f1} show the scatter plots of galaxy magnitudes: 
K vs. r (left panel), u vs. r (middle panel),  and K vs u (right panel). 
Using a linear fit for K vs. r, we determined the r-band magnitude 
limit corresponding to $K_{\rm lim}=13.57$, which led us to $r_{\rm lim}=16.54$. 
To determine the limit in the u-band, 
we fit the distribution u vs. r\footnote{Instead of using the 'u vs. K' distribution, 
we chose using 'u vs. r' to make the linear fit since it has smaller dispersion}. 
For $r_{\rm lim}=16.54$, we found $u_{\rm lim}=18.73$. 
Given the spread of the distributions around the linear fits, 
adopting a fixed magnitude limit in one or another band inevitably 
leads to a different galaxy sampling towards the fainter magnitudes. 
This is a fact that is also present in the observational catalogues 
limited by flux, as many of the catalogues used in the literature to
 identify CGs in different bands. 
We will also explore how this affect the samples of CGs 
extracted from each catalogue.
The number of galaxies comprised in each mock catalogue are 
quoted in Table~\ref{tab1}.

\begin{table}
\scriptsize
\centering
\caption{Group properties for the total and restricted samples of CGs\label{tab2}}
\tabcolsep 1.0pt
\begin{tabular}{|c|r|r|r|r|r|r|}
\hline
& \multicolumn{3}{c|}{Total Samples}&\multicolumn{3}{c|}{Restricted Samples}\\
\cline{2-7}
Sample       & K   & r   & u & $\rm \widetilde{K}$ & $\tilde{\rm r}$ & $\tilde{\rm u}$\\
\# CG   & 447 & 406 & 276 & 382 & 289 & 231 \\
\hline
                          $\theta_G$ & $ 3.2 \ ( 0.2)$ & $ 3.4 \ ( 0.2)$ & $ 3.5 \ ( 0.3) $ & $ 3.3 \ ( 0.2)$ & $ 3.7 \ ( 0.2)$ & $ 3.5 \ ( 0.3) $\\
                               $r_p$ & $  67 \ (   3)$ & $  72 \ (   3)$ & $  66 \ (   4) $ & $  65 \ (   3)$ & $  73 \ (   4)$ & $  62 \ (   4) $\\
                       $R_{\rm vir}$ & $  90 \ (   5)$ & $ 106 \ (   5)$ & $  96 \ (   7) $ & $  86 \ (   5)$ & $ 102 \ (   6)$ & $  93 \ (   7) $\\
             $\langle d_{ij}\rangle$ & $  73 \ (   4)$ & $  82 \ (   4)$ & $  73 \ (   5) $ & $  72 \ (   4)$ & $  81 \ (   4)$ & $  70 \ (   4) $\\
                           $s_\perp$ & $ 105 \ (   6)$ & $ 117 \ (   6)$ & $ 113 \ (   8) $ & $ 102 \ (   6)$ & $ 114 \ (   8)$ & $ 101 \ (   7) $\\
                       $s_\parallel$ & $ 103 \ (  14)$ & $ 119 \ (  17)$ & $ 156 \ (  31) $ & $  99 \ (  14)$ & $ 117 \ (  19)$ & $ 140 \ (  28) $\\
                               $s_4$ & $ 168 \ (  15)$ & $ 182 \ (  16)$ & $ 205 \ (  27) $ & $ 162 \ (  15)$ & $ 180 \ (  19)$ & $ 187 \ (  24) $\\
                          $\sigma_v$ & $ 280 \ (  14)$ & $ 281 \ (  15)$ & $ 253 \ (  21) $ & $ 270 \ (  15)$ & $ 284 \ (  18)$ & $ 265 \ (  22) $\\
                 $H_0 \, t_{\rm cr}$ & $ 2.7 \ ( 0.2)$ & $ 2.9 \ ( 0.3)$ & $ 2.9 \ ( 0.4) $ & $ 2.7 \ ( 0.2)$ & $ 2.8 \ ( 0.3)$ & $ 2.6 \ ( 0.3) $\\
                ${\cal M}_{\rm vir}$ & $ 6.9 \ ( 0.9)$ & $ 7.3 \ ( 1.1)$ & $ 6.5 \ ( 1.4) $ & $ 6.3 \ ( 0.9)$ & $ 7.8 \ ( 1.4)$ & $ 6.5 \ ( 1.4) $\\
                       $K_{\rm bri}$ & $10.0 \ ( 0.1)$ & $10.2 \ ( 0.1)$ & $ 9.9 \ ( 0.1) $ & $ 9.9 \ ( 0.1)$ & $ 9.9 \ ( 0.1)$ & $ 9.8 \ ( 0.1) $\\
                       $r_{\rm bri}$ & $12.8 \ ( 0.1)$ & $13.0 \ ( 0.1)$ & $12.6 \ ( 0.1) $ & $12.7 \ ( 0.1)$ & $12.7 \ ( 0.1)$ & $12.6 \ ( 0.1) $\\
                       $u_{\rm bri}$ & $15.2 \ ( 0.1)$ & $15.3 \ ( 0.1)$ & $15.0 \ ( 0.1) $ & $15.1 \ ( 0.1)$ & $15.0 \ ( 0.1)$ & $14.9 \ ( 0.1) $\\
        \scriptsize $\Delta K_{12}$  & $ 1.3 \ ( 0.1)$ & $ 1.2 \ ( 0.1)$ & $ 1.1 \ ( 0.1) $ & $ 1.3 \ ( 0.1)$ & $ 1.2 \ ( 0.1)$ & $ 1.1 \ ( 0.1) $\\
        \scriptsize $\Delta r_{12}$  & $ 1.3 \ ( 0.1)$ & $ 1.2 \ ( 0.1)$ & $ 1.0 \ ( 0.1) $ & $ 1.3 \ ( 0.1)$ & $ 1.2 \ ( 0.1)$ & $ 1.1 \ ( 0.1) $\\
        \scriptsize $\Delta u_{12}$  & $ 1.3 \ ( 0.1)$ & $ 1.2 \ ( 0.1)$ & $ 1.1 \ ( 0.1) $ & $ 1.3 \ ( 0.1)$ & $ 1.2 \ ( 0.1)$ & $ 1.1 \ ( 0.1) $\\
                             $\mu_K$ & $21.9 \ ( 0.1)$ & $22.2 \ ( 0.1)$ & $22.1 \ ( 0.1) $ & $21.9 \ ( 0.1)$ & $22.2 \ ( 0.1)$ & $21.8 \ ( 0.1) $\\
                             $\mu_r$ & $24.8 \ ( 0.1)$ & $25.1 \ ( 0.1)$ & $24.9 \ ( 0.1) $ & $24.7 \ ( 0.1)$ & $25.1 \ ( 0.1)$ & $24.7 \ ( 0.1) $\\
                             $\mu_u$ & $27.1 \ ( 0.1)$ & $27.4 \ ( 0.1)$ & $27.2 \ ( 0.1) $ & $27.1 \ ( 0.1)$ & $27.4 \ ( 0.1)$ & $27.0 \ ( 0.1) $\\
                               $L_K$ & $ 231 \ (  14)$ & $ 219 \ (  15)$ & $ 194 \ (  18) $ & $ 227 \ (  14)$ & $ 224 \ (  19)$ & $ 207 \ (  18) $\\
                               $L_r$ & $  68 \ (   4)$ & $  66 \ (   4)$ & $  59 \ (   4) $ & $  67 \ (   4)$ & $  67 \ (   5)$ & $  61 \ (   5) $\\
                               $L_u$ & $  41 \ (   3)$ & $  41 \ (   3)$ & $  44 \ (   4) $ & $  41 \ (   3)$ & $  42 \ (   4)$ & $  41 \ (   3) $\\
\scriptsize ${\cal M}_{\rm v}$/$L_K$ & $  29 \ (   3)$ & $  32 \ (   4)$ & $  32 \ (   6) $ & $  27 \ (   3)$ & $  30 \ (   5)$ & $  30 \ (   5) $\\
\scriptsize ${\cal M}_{\rm v}$/$L_r$ & $ 100 \ (  11)$ & $ 109 \ (  14)$ & $ 106 \ (  19) $ & $  94 \ (  12)$ & $ 104 \ (  17)$ & $ 103 \ (  19) $\\
\scriptsize ${\cal M}_{\rm v}$/$L_u$ & $ 158 \ (  20)$ & $ 178 \ (  24)$ & $ 161 \ (  27) $ & $ 136 \ (  20)$ & $ 174 \ (  30)$ & $ 167 \ (  30) $\\
                              $T1_K$ & $0.56 \ (0.03)$ & $0.56 \ (0.03)$ & $0.75 \ (0.04) $ & $0.54 \ (0.02)$ & $0.54 \ (0.03)$ & $0.65 \ (0.04) $\\
                              $T2_K$ & $0.61 \ (0.02)$ & $0.61 \ (0.02)$ & $0.68 \ (0.03) $ & $0.61 \ (0.02)$ & $0.61 \ (0.03)$ & $0.65 \ (0.03) $\\
                              $T1_r$ & $0.56 \ (0.02)$ & $0.57 \ (0.02)$ & $0.72 \ (0.04) $ & $0.54 \ (0.02)$ & $0.56 \ (0.03)$ & $0.64 \ (0.03) $\\
                              $T2_r$ & $0.61 \ (0.02)$ & $0.61 \ (0.02)$ & $0.67 \ (0.03) $ & $0.59 \ (0.02)$ & $0.60 \ (0.03)$ & $0.63 \ (0.03) $\\
                              $T1_u$ & $0.63 \ (0.03)$ & $0.66 \ (0.03)$ & $0.71 \ (0.04) $ & $0.61 \ (0.03)$ & $0.65 \ (0.03)$ & $0.68 \ (0.04) $\\
                              $T2_u$ & $0.67 \ (0.02)$ & $0.71 \ (0.03)$ & $0.69 \ (0.03) $ & $0.66 \ (0.03)$ & $0.70 \ (0.03)$ & $0.66 \ (0.03) $\\
 \% Reals & 55.6 & 52.0 & 44.4 & 57.9 & 52.2 & 48.9   \\
\hline 
\end{tabular}
\parbox{8cm}{
\small
{\ct {\bf Notes:} In each cell, the format $\rm xx \ (ss)$ contains the median ($\rm xx$) 
and the shift ($\rm ss$) to construct an approximated 95\% confidence interval, 
$CI = xx \pm ss$ (see text for details),
except for the $T_1$ and $T_2$ values where the quantity in parentheses
are the error bars computed using the bootstrap resampling technique. 
Units: $\theta_G$ = $\rm arcmin$; $r_p$, $R_{\rm vir}$, $\langle d_{\rm ij} \rangle$, 
$s_\perp$, $s_\parallel$ and $s_4$ = $\rm kpc \ h^{-1}$; $\sigma_v$ = $\rm km \ s^{-1}$ 
; $H_0 \, t_{\rm cr}$ = $10^{-2}$ ; 
${\cal M}_{\rm vir}$ = $10^{12} \ {\cal M}_{\odot} \ \rm h^{-1}$ ;
magnitude gaps are calculated in absolute magnitudes 
for each photometric band; $\mu$ = $\rm mag/arcsec^2$; $L$ = $10^{9} \ L_{\odot} \ \rm h^{-2}$; ${\cal M}_{\rm vir}/L$ = $ {\rm h} \, {\cal M}_{\odot}/L_{\odot} $.}
}
\end{table}

\subsection{The compact group identification}
\label{iden}
We identified mock CGs in the galaxy lightcones using the criteria defined 
by \cite{Diaz_Gimenez+12}: 
\begin{itemize}
\item Population: $4 \le N \le 10$
\item Compactness: $\mu \le \mu_{\rm lim}$ [mag/arcsec$^{-2}$]
\item Isolation: $\Theta_N > 3 \Theta_G$
\item Flux limit: $m_{\rm bri} \le m_{\rm lim} - 3$
\item Velocity filtering: $|v_i - <v>| \le 1000 \, \rm km \, s^{-1}$
\end{itemize}
where N is the number of galaxies whose K-band magnitudes are within a 
3-magnitude range from the brightest galaxy, 
$m_{\rm bri}$ is the apparent magnitude of the brightest galaxy of 
the group; $\mu$ is the mean surface brightness in a given band, averaged over 
the smallest circle that circumscribes the galaxy centres; $\Theta_G$ is 
the angular diameter of the smallest circumscribed circle; $\Theta_N$ is 
the angular diameter of the largest
concentric circle that contains no other galaxies within the considered 
magnitude range or brighter; $v_i$ is the radial velocity of each galaxy 
member and $<v>$ is the median of the radial velocity of the members. 
In Table~\ref{tab1} we show the limits adopted in each band 
for the compactness and flux limit criteria. 
In the K-band, we adopted the values described in \cite{Diaz_Gimenez+12}. 
To determine the limiting value for the compactness criterion 
in the other two bands, following \cite{Diaz_Gimenez+12}, 
we examined the resulting sample of CGs identified 
in the K-band and determined the mean band shifts: $(r-K)^0=2.73$ and $(u-K)^0=5.05$.
These are shown with solid lines in the bottom panels of Fig~\ref{f1}.
Using these values we shifted the corresponding $\mu$ value in the K band to obtain
the corresponding values in the other two bands.
We also checked for CGs embedded within larger CGs that also meet the criteria described above.
Following \cite{Diaz_Gimenez+10}, for such groups, we kept the larger group 
(for a complete description of the algorithm, see Figure 1 in that paper). 

We have also considered the fact that galaxies in the mock catalogues are 
just point-sized particles, therefore we have included the blending of 
galaxies in projection on the plane of the sky which modify the number of 
detectable objects, changing the population of the compact groups. 
According to the morphological type of each galaxy - determined based 
on the ratio of stellar mass of the bulge and the total stellar mass 
\citep{Bertone_2007} -, we computed their half-light radii in each band 
as a function of the stellar mass of each mock galaxy following the 
prescriptions of \cite{Lange_2015} (see Table~\ref{tab1}). 
Finally, we considered two galaxies 
as blended if the angular separation between the two galaxies is smaller 
than the sum of their angular half-light radii.

Within a solid angle of 4$\pi$, we identified $447$ CGs in the K-band (K-CGs), 
$406$ in the r-band (r-CGs), and $276$ in the u-band (u-CGs). 

\subsection{Compact group properties}
We measured several group properties that will be used to compare the three CG samples. 
These properties are:
\begin{itemize}
\item $\theta_G$: angular diameter of the minimum circle that encloses all the group members
\item $r_p$: projected radius of the minimum circle
\item $R_{\rm vir}$: projected virial radius of the group computed as $R_{\rm vir}= 2 N (N-1) (\sum_{ij} \frac{1}{d_{ij}})^{-1}$
where  $d_{ij}$ are the projected separations between galaxies.  
\item $\langle d_{ij}\rangle$: median of the projected intergalaxy separations 
\item $s_\perp$: maximum projected separation between the four closest galaxies			
\item $s_\parallel$: maximum comoving line-of-sight separation between 
the four closest galaxies.  
\item $s_4$: 3D comoving maximum interparticle separation between the four closest galaxies
\item $\sigma_v$: group gapper radial velocity dispersion
\item $H_0 \, t_{\rm cr}$: dimensionless crossing time computed as $H_0 \, t_{\rm cr}=
\frac{\pi\x100 \ {\rm h}}{2\sqrt{3}}\frac{\langle d_{ij}\rangle}{\sigma_v}$
\item ${\cal M}_{\rm vir}$: virial mass computed as ${\cal M}_{\rm vir}= \frac{3\pi}{2 G} R_{\rm vir} \, \sigma_v^2 $  
\item $K_{\rm bri}$, $r_{\rm bri}$, $u_{\rm bri}$: observer frame apparent magnitude 
of the brightest galaxy in the three bands 
\item $\Delta K_{12}$, $\Delta r_{12}$, $\Delta u_{12}$: rest frame absolute 
magnitude difference between the first- and the second-ranked galaxies 
\item $\mu_K$, $\mu_r$, $\mu_u$: group surface brightness in the three bands 
\item $L_K$, $L_r$, $L_u$: total group luminosity in the three bands 	
\item ${\cal M}_{\rm v}/L_K$, ${\cal M}_{\rm v}/L_r$, ${\cal M}_{\rm v}/L_u$: mass-to-light ratio in the three bands
\end{itemize}			

We also computed for each sample and in each band, the Tremaine-Richstone 
statistics, T1 and T2 \citep{Tremaine+77}: 
\begin{equation}
\nonumber
T1=\frac{\sigma(M_1)}{<M_2-M_1>}, \ \ \ T2=\frac{1}{\sqrt{0.677}} \frac{\sigma(M_2-M_1)}{<M_2-M_1>}
\end{equation}
Groups with a first-ranked galaxy much brighter than the second-ranked galaxy exhibit values of T1 and T2 lower than unity. 
According to \cite{mamon87}, mergers within groups reduce the values of T1 and T2 below 0.7.
Finally, we split the samples of CGs into physically dense (Reals) and 
chance alignments (CAs) following the 3D classification performed 
by \cite{Diaz_Gimenez+10} which involves $s_4$, $s_\perp$ and $s_\parallel$. 

The median of these properties and the percentages of Reals CGs are quoted in Table~\ref{tab2}. 
The shifts to compute the 95\% confidence interval (CI) for the median are quoted within 
 parentheses, and are given by $1.58 \times IQR/\sqrt{n}$, where $n$ is the number of objects in 
the sample and $IQR$ is the interquartile range \citep{boxplot14}. 

Using the non-parametric Kolmogorov-Smirnov test, we measured the probability that two
samples are drawn from the same distribution. In Table~\ref{tab3}, we
quote the p-values obtained from the test when comparing pairs of samples. 
We have highlighted the cells when the probability is lower than a typical critical 
value of 0.05, which indicates statistical differences for the distributions of a given 
property between the two samples.
\begin{table}
\centering
\scriptsize
\caption{P-values of the Kolmogorov-Smirnov two sample test. Each column combines different
pairs of CG samples: the first three columns compare the total samples, 
the following three columns compare the restricted samples, while the last three columns compare
the total vs restricted samples in each band. 
We highlighted in grey the cells with p-values lower than 0.05.\label{tab3}}
\tabcolsep 0.8pt
\begin{tabular}{|c|c|c|c|c|c|c|c|c|c|}
\hline
 & K - r & K - u & u - r & $\rm \widetilde{K}$ - $\rm \tilde{r}$ & $\rm \widetilde{K}$ - $\rm \tilde{u}$ & $\rm \tilde{u}$ - $\rm \tilde{r}$ & K - $\rm \widetilde{K}$  & r - $\rm \tilde{r}$ & u - $\rm \tilde{u}$\\
\hline
                     $\theta_G$ &$   0.16  $		  &$   0.08  $		    &$   0.32  		     $&$ \ct 4\x 10^{-3}    $&$              0.72  $&$              0.19  $&$              1.00  $&$              0.11  $&$              1.00  $\\
                          $r_p$ &$\ct 0.02  	$	  &$   0.99  $		    &$   0.13  		     $&$ \ct 7\x 10^{-3}    $&$              0.56  $&$ \ct 1\x 10^{-3}    $&$              1.00  $&$              1.00  $&$              0.51  $\\
                  $R_{\rm vir}$ &$   \ct 1\x10^{-4}      $&$   0.06  $		    &$   0.19  		     $&$ \ct 3\x 10^{-5}    $&$              0.18  $&$              0.05  $&$              0.97  $&$              1.00  $&$              0.66  $\\
               $\langle d_{ij}\rangle$ &$   \ct 0.01  $	  &$   0.42  $		    &$   0.12  		     $&$ \ct 2\x 10^{-3}    $&$              0.47  $&$ \ct 4\x 10^{-3}    $&$              0.98  $&$              1.00  $&$              0.31  $\\
$s_\perp$ 			&$   \ct 0.02  $	  &$   0.23  $		    &$   0.87  		     $&$              0.04  $&$              0.57  $&$              0.14  $&$              0.99  $&$              1.00  $&$              0.29  $\\ 
$s_\parallel$ 			&$   0.56  $		  &$   \ct 5\x10^{-4}      $&$   \ct 0.02  	     $&$              0.68  $&$ \ct 9\x 10^{-3}    $&$              0.26  $&$              1.00  $&$              1.00  $&$              0.79  $\\ 
                          $s_4$ &$   0.26  $		  &$   \ct 3\x10^{-3}      $&$   0.06  		     $&$              0.24  $&$              0.10  $&$              0.75  $&$              1.00  $&$              1.00  $&$              0.68  $\\ 
                     $\sigma_v$ &$   1.00  $		  &$   0.23  $		    &$   0.24  		     $&$              0.72  $&$              0.94  $&$              0.66  $&$              0.99  $&$              1.00  $&$              1.00  $\\ 
 $H_0 \, t_{\rm cr}$ 		&$   0.19  $		  &$   0.09  $		    &$   0.88  		     $&$              0.41  $&$              0.96  $&$              0.36  $&$              1.00  $&$              1.00  $&$              0.55  $\\
${\cal M}_{\rm vir}$ 		&$   0.45  $		  &$   0.40  $		    &$   0.27  		     $&$              0.08  $&$              0.78  $&$              0.70  $&$              0.91  $&$              1.00  $&$              1.00  $\\ 
                  $K_{\rm bri}$ &$   \ct 1\x10^{-5}      $&$   \ct 0.01  $	    &$   \ct 1\x10^{-4}      $&$              0.93  $&$              0.29  $&$              0.26  $&$              0.10  $&$ \ct 1\x 10^{-7}    $&$    \ct       0.03  $\\ 
                  $r_{\rm bri}$ &$   \ct 1\x10^{-4}      $&$   0.37  $		    &$   \ct 1\x10^{-4}      $&$              0.95  $&$              0.25  $&$              0.37  $&$ \ct          0.03  $&$ \ct 1\x 10^{-8}    $&$              0.06  $\\ 
                  $u_{\rm bri}$ &$   \ct 8\x10^{-3}      $&$   \ct 2\x10^{-3}      $&$   \ct 1\x10^{-5}      $&$              0.99  $&$              0.20  $&$              0.60  $& $\ct 0.01           $&$ \ct 6\x 10^{-8}    $&$              1.00  $\\ 
          $\Delta K_{12}$       &$   \ct 0.04  $	  &$   \ct 1\x10^{-3}      $&$   0.18  		     $&$              0.24  $&$   \ct        0.03  $&$              0.71  $&$              1.00  $&$              0.96  $&$              0.96  $\\ 
          $\Delta r_{12}$       &$   0.08  $		  &$   \ct 1\x10^{-3}      $&$   0.25  		     $&$              0.35  $&$   \ct        0.01  $&$              0.64  $&$              1.00  $&$              1.00  $&$              1.00  $\\ 
          $\Delta u_{12}$       &$   0.11  $		  &$   \ct 0.03  $	    &$   0.99  		     $&$              0.35  $&$   \ct        0.01  $&$              0.64  $&$              1.00  $&$              1.00  $&$              1.00  $\\ 
                        $\mu_K$ &$   \ct 4\x10^{-3}      $&$   0.10  $		    &$   0.20  		     $&$ \ct 7\x 10^{-3}    $&$              0.72  $&$ \ct 1\x 10^{-3}    $&$              0.97  $&$              1.00  $&$              0.19  $\\ 
                        $\mu_r$ &$   \ct 4\x10^{-3}      $&$   0.19  $		    &$   0.13  		     $&$ \ct 5\x 10^{-3}    $&$              0.39  $&$ \ct 1\x 10^{-3}    $&$              0.94  $&$              1.00  $&$              0.23  $\\
                        $\mu_u$ &$   \ct 0.01  $	  &$   0.50  $		    &$   \ct 0.04  	     $&$ \ct          0.01  $&$              0.26  $&$ \ct 1\x 10^{-3}    $&$              0.86  $&$              1.00  $&$              0.59  $\\ 
          $L_K $ 		&$   0.73  $		  &$   \ct 8\x10^{-3}  $&$   0.14  		     $&$              0.87  $&$              0.24  $&$              0.56  $&$              1.00  $&$              1.00  $&$              0.75  $\\ 
          $L_r $ 		&$   0.75  $		  &$   \ct 0.03  $	    &$   0.17  		     $&$              0.89  $&$              0.50  $&$              0.51  $&$              1.00  $&$              1.00  $&$              1.00  $\\ 
          $L_u $ 		&$   1.00  $		  &$   0.79  $		    &$   0.73  		     $&$              0.79  $&$              0.99  $&$              0.80  $&$              1.00  $&$              1.00  $&$              0.96  $\\ 
\tiny  ${\cal M}_{\rm v}$/$L_K$ &$   0.29  $		  &$   0.11  $		    &$   0.76  		     $&$              0.10  $&$              0.19  $&$              0.97  $&$              0.99  $&$              1.00  $&$              0.97  $\\ 
\tiny  ${\cal M}_{\rm v}$/$L_r$ &$   0.34  $		  &$   0.20  $		    &$   0.91  		     $&$              0.12  $&$              0.29  $&$              1.00  $&$              0.99  $&$              1.00  $&$              1.00  $\\ 
\tiny  ${\cal M}_{\rm v}$/$L_u$ &$   0.39  $		  &$   0.73  $		    &$   0.51  		     $&$              0.21  $&$              0.48  $&$              0.91  $&$              0.97  $&$              1.00  $&$              1.00  $\\ 
\hline
\end{tabular}
\end{table}

Analysing the values of the medians (Table~\ref{tab2}) and 
the comparison between the properties in different photometric 
bands (Table~\ref{tab3}) we found: 
\begin{itemize}
\item K-CGs have projected sizes ($r_p$,$R_{\rm vir}$, $d_{ij}$) 
statistically smaller than r-CGs.
\item K-CGs show smaller 3D interparticle separation ($s_4$) than the u-CGs. 
\item u-CGs present the lowest velocity dispersion (although the difference is not significant)
\item the magnitude gap in the three bands is larger for the K-CGs than for the u-CGs.
\item K-CGs show greater compactness (lower surface brightness) in the three bands 
than the r-CGs, and similar to the u-CGs (except for $\mu_u$). 
\item the luminosity in the u-band is similar for CGs identified in any of the three bands. 
However, the K-band and r-band luminosities of the K-CGs are higher than the luminosities 
of the u-CGs.  
\item the brightest galaxies of the u-CGs are brighter than the brightest galaxies of the 
K- or r-CGs in the three bands. 
\item The $T_1$ and $T_2$ computed in the K- and r- band show very low values for the
K-CGs and r-CGs samples which means very different first and second-ranked galaxies
in luminosity, indicating clear signals of mergers within groups. Even though the u-CGs 
show $T_1$ and $T_2$ below the unity, their values are close to 0.7.
\item the K-CG sample comprises more Real CGs than the observed in the other photometric
bands. The lowest value for the percentage of Real CGs is obtained in the u-CG sample.
\end{itemize}

\cite{Diaz_Gimenez+12} and \cite{herfer15} compared CGs identified in different bands 
and with different algorithms. Regarding the projected sizes, both studies agreed that 
groups identified in the R- \citep{hickson92,allam00} or r- band \citep{mcconnachie09} where 
smaller in projection than groups identified in the K-band \citep{Diaz_Gimenez+12}, 
and \citeauthor{herfer15} also stated that their UV-CGs are the smallest.  
In this work we found a different result: K-CGs and u-CGs are smaller than r-CGs. 
This controversial result might be expected since the difference 
in the algorithms used to identify the observational 
samples that have been compared introduce the differences in sizes reported in the literature. 
Using friends-of-friends algorithms, or Hickson-like criteria based on the projected sizes, and also 
including/excluding the CG-in-CG options introduces a bias in the projected sizes of the resulting groups.
In this work, we avoided this issue by using the same algorithm in the three photometric bands, therefore the
results reported here are only due to intrinsic differences between bands.
Also, \cite{Diaz_Gimenez+12} and \cite{herfer15} found that the crossing times of groups identified in 
the B and UV bands are larger than K-CGs which in turn are larger than R- or r-CGs. 
Here, we do not find significant differences between the crossing times of groups identified in the 
three bands. 
Moreover, \cite{Diaz_Gimenez+12} found that the mean projected interparticle separation between galaxies 
in R-CGs is smaller than in K-CGs, 
which are smaller than B-CGs. However, \cite{herfer15} found that the smallest
interparticle separation was in the UV-CGs, with the r-CGs presenting the largest separations. 
In this work we found the r-CGs with the largest interparticle separation in agreement with \cite{herfer15}.
Finally, \cite{Diaz_Gimenez+12} found that the only sample showing T1 and T2 values below unity was 
the one identified in the K-band. On the contrary, in this work using the same algorithm in the three bands, 
we do find that all the values are below unity, being higher in the u-band and very similar in the K- and r-bands.

%


\begin{figure}
\begin{center}
\includegraphics[width=8.5cm]{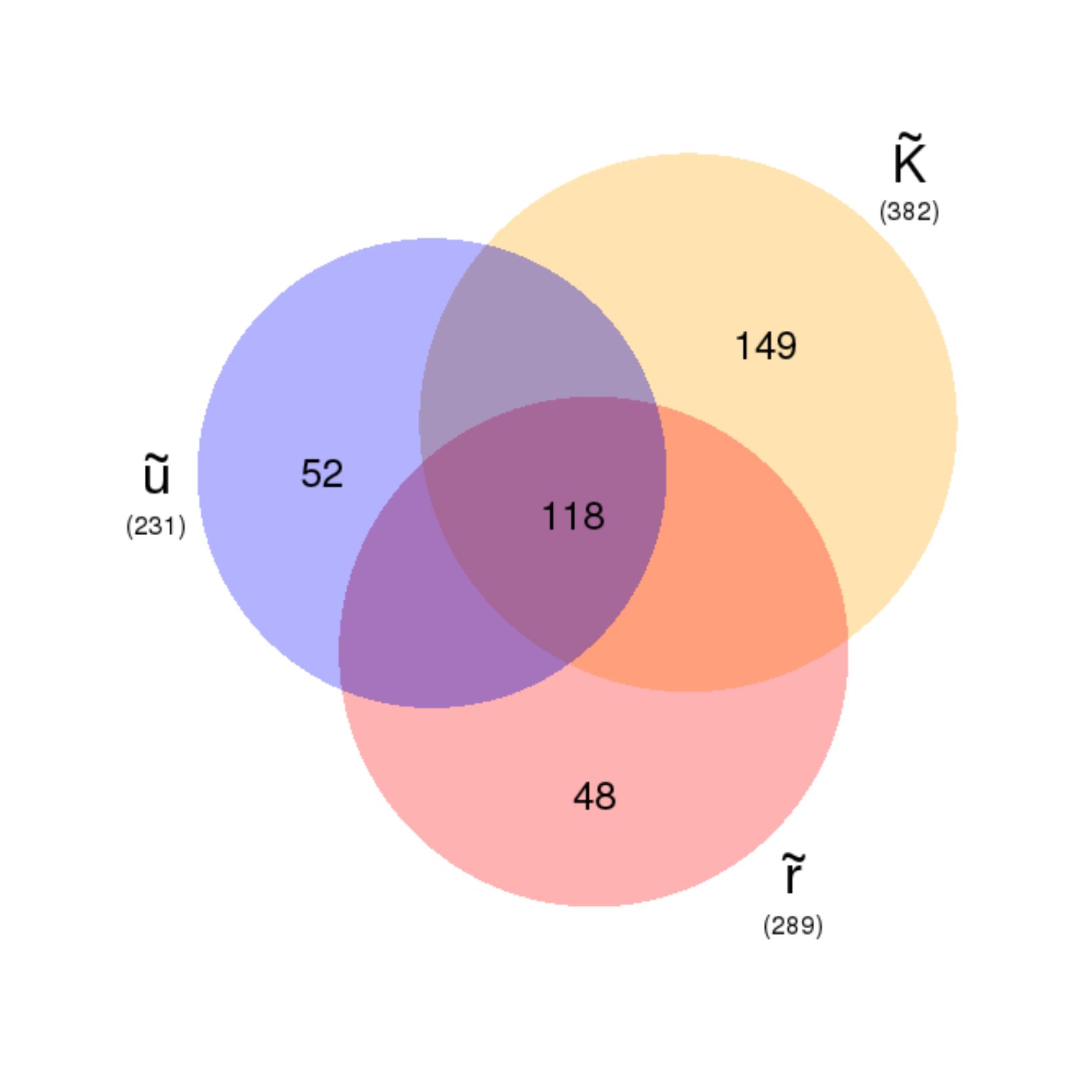}
\caption{\label{Venn} Schematic Venn diagrams that shows the different 
overlapping regions that are obtained when comparing the group identified on each
photometric band. The number of groups in the total restricted samples are quoted in parentheses.
The number of groups identified only in a given band and those that are common in the three bands are also 
quoted inside the coloured circles.}
\end{center}
\end{figure}
\begin{figure*}
\begin{center}
\includegraphics[width=8.8cm]{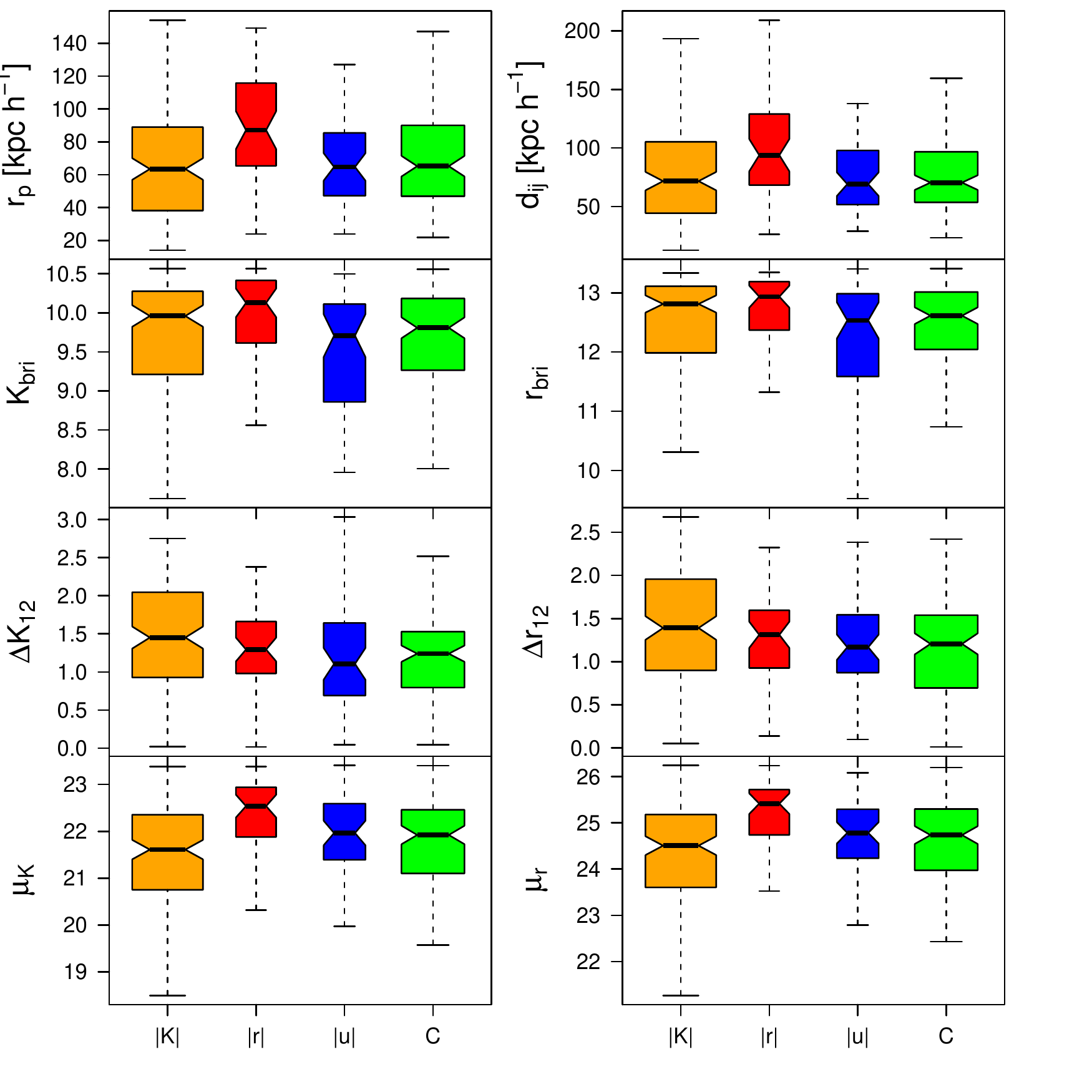}
\hskip -0.5cm
\includegraphics[width=8.8cm]{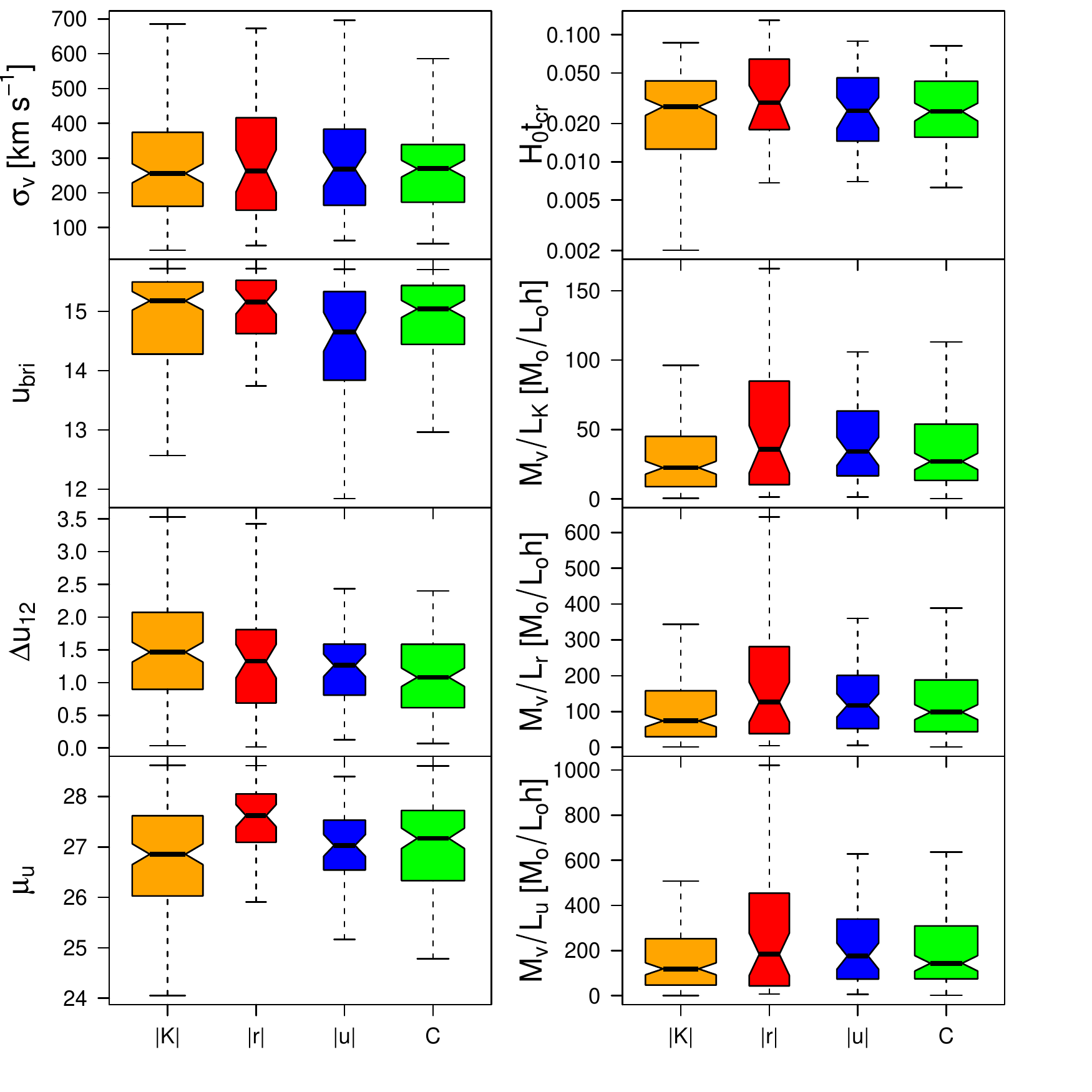}
\caption{\label{puros} Boxplot diagrams of some of the properties for the pure and common CGs.
The notches indicates the approximate 95\% confidence interval for the medians, while
the widths are proportional to the square roots of the number of CGs in each sample.
The complete list of values for the medians and their corresponding 95\% confidence
intervals for all the properties studied in this work are quoted in Table~\ref{pure_mix}.
}
\end{center}
\end{figure*}

\subsection{Restricted samples}

One may wonder whether the differences we found in the previous section 
could be biased due to the fact that each sample has been selected with 
different criteria depending on the band ($\mu_{\rm lim}$ and $m_{\rm bri_{lim}}$). 
Therefore, in this section we
normalised the criteria to avoid any dependence on the selection function.

We defined the ``restricted'' samples 
($\rm \widetilde{K}$-, $\rm \tilde{r}$- and $\rm \tilde{u}$-CGs) 
by selecting CGs in each catalogue that also satisfy 
the compactness and flux criteria in the other two bands. 
For instance, the $\rm \widetilde{K}$-CGs are a subsample of the K-CGs that
also satisfy that $\mu_{\rm r} < 26.33$, and $\mu_{\rm u} < 28.65$, and 
$r_{\rm brightest} < 13.54$, and $u_{\rm brightest} < 15.73$. 
We selected the $\rm \tilde{r}$- and $\rm \tilde{u}$-CGs in a similar way.

We found that 85.5\% of the K-CGs survive the triple restriction, 
while 71.2\% of the r-CGs and 83.7\% of the u-CGs conform their restricted samples. 
In Table~\ref{tab2} we quoted the medians of the properties of these restricted samples 
and the quantities to compute their 95\% confidence interval, while 
in Table~\ref{tab3} we quoted the p-values obtained from the KS-test used for 
the comparison between samples. 

In general, the restricted samples only differ from the original 
samples in those properties that are explicitly dependent on 
the limit adopted on the magnitudes (see the last three columns of Table~\ref{tab3}). 
As a consequence, the differences we found between the original samples
in the magnitude of the brightest galaxies and total luminosities of the groups are 
not longer present once we imposed the same limits for the three catalogues.

However, the differences in sizes, magnitude gaps and surface brightness remain in 
the restricted samples: \~u-CGs are smallest in projection, the K-CGs have the largest magnitude 
gap (although non-significant compared to \~r-CGs), while the \~r-CGs have the largest surface 
brightness -lower compactness- and this result is 
more significant now using the restricted samples.

As a by product, having restricted the samples in the three bands at the same time has 
slightly increased the probability of selecting physically dense groups (Reals) 
in the K and u bands, which is something desirable in an observational 
sample where we cannot easily distinguish between chance alignments 
and physically dense groups.


\section{Cross-identified compact groups: pure and common identifications}
\label{purmix}
Although the CGs belonging to the restricted samples accomplish the 
compactness and flux criteria in the three bands at the same time, 
those groups are not the same. 
In Fig.~\ref{Venn} we show the Venn diagram of the three sets of restricted groups. 
In this section we are interested in analysing the CGs that can be identified in the three
bands at the same time and those CGs that can be identified only in one particular band. 

To select CGs that were identified in the three bands, 
we first looked for CGs that were identified in two bands at the same time, 
and then we cross-checked if those groups were also in the third band. 
When comparing CGs in two bands, there could be CGs that belonging to both catalogues
are not necessarily exactly the same: there could be a slightly different number of members 
in the two bands by including/excluding a galaxy in one or the other band. 
Therefore, instead of making a member-to-member comparison, 
we adopted a simpler criterion to determine whether a CG that belongs to two 
different catalogues is the same: if the angular distance between the centres of the minimum 
circles is less than twice the angular radius of the CG and 
the difference in radial velocity of the centres is less than $1000 \, \rm km \, s^{-1}$. 
Using this criterion we found $118$ CGs in common in the three bands (C-CGs). 
They represent $31\%$ of the $\rm \widetilde{K}$-CGs, $41\%$ of the $\rm \tilde{r}$-CGs 
and $51\%$ of the $\rm \tilde{u}$-CGs. 

On the other hand, CGs in each band that have not a counterpart in any of the other two 
catalogues conform the sample of ``pure'' CGs ($|K|$-,$|r|$-,$|u|$-CGs). 
We found that $\sim 39\%$ of the $\rm \widetilde{K}$-CGs can only be identified 
when analysing a sample of galaxies selected in the K-band, 
$\sim 17\%$ of the $\rm \tilde{r}$-CGs exist only in the r-band, while $\sim 23\%$ 
of the $\rm \tilde{u}$-CGs exist only in the u-band. 

We show a comparison of some of the properties of CGs belonging to each of these subsamples 
using boxplot diagrams in Fig.~\ref{puros}.
In these diagrams, the notches correspond to the approximated 95\% confidence intervals. 
In general, when notches do not overlap, the medians can be judged to differ 
significantly \citep{boxplot14}, but overlap does not rule out a significant difference.
In appendix~\ref{apen}, we show the values of the medians and their confidence intervals for all the properties
used in this work for groups in each of these samples, as well as the p-values of the KS-tests 
between pairs of samples (Tables~\ref{pure_mix} and \ref{KS_pure_mix}, respectively).

From the analysis of the properties (see Fig.~\ref{puros} and Tables~\ref{pure_mix}) 
for the different CG samples in different photometric bands we can mention the following highlights: 
\begin{itemize}
\item $|K|$-CGs present the smallest projected virial radius, virial masses, surface brightness 
(highest compactness), mass-to-light ratios and the largest magnitude gaps, 
even larger than all the samples previously analysed. 
They also produce the lowest values of T1 and T2.
\item $|r|$-CGs present the largest projected sizes, virial masses, surface brightness 
(lowest compactness), and mass-to-light ratios. 
\item $|u|$-CGs present 
the brightest first ranked galaxy, 
and the largest 3D and line-of-sight interparticle separations (which makes the $|u|$ sample to present 
the lowest percentage of physically dense groups - according to our criterion based on the 3D comoving separations). On average, these groups show one of the largest values of T1 and T2.
\item $C$-CGs present the smallest 
3D, line-of-sight and projected comoving sizes of the four closest neighbours
(which implies having the largest fraction of Real CGs).
\end{itemize}

By comparing Table~\ref{KS_pure_mix} and Table~\ref{tab3}, we can analyse the
effect of including/excluding groups that exist in one or other band. 
In general, the differences found between restricted CGs remain, but also 
a few other differences arise when comparing pure vs pure samples that were 
not present in the restricted samples.  We found that 
$|K|$ virial masses are significantly lower than $|r|$ virial masses, and this is also 
observed when analysing the mass-to-light ratios. 
In addition, the 3D comoving interparticle separations of $|u|$-CGs are statistically larger 
than for the $|K|$-CGs. 
Also, the brightest galaxies (in any band) of the $|r|$-CGs are typically brighter than 
the brightest galaxies in $|u|$-CGs.

Finally, when comparing the pure CGs with the common CGs we found a couple of differences with 
one or other sample. 
The projected sizes of C-CGs are quite similar to the $|u|$, however, 
the 3D interparticle separations of 
the C-CGs are significantly smaller than the $|u|$. 
The magnitude gaps in the three bands of the $|K|$ sample
are typically larger than the gaps for the common groups, 
meaning the common CGs are conformed by 
more similar neighbour galaxies than the pure K. 
The $|r|$-CGs are statistically less ``compact'' (fainter $\mu$) 
than the common CGs. Also, the brightest galaxies in C-CGs are brighter than 
their counterpart in $|r|$-CGs.
Interestingly, the common CGs present the largest percentage of Real CGs. 
This is a novel result that may help selecting observational samples with the 
lowest percentage of chance alignments, although it is necessary to have data 
in multiple photometric bands. 
Some of the differences that we found between the samples of ``pure'' 
CGs are not observed in the restricted 
samples given the existence of the common CGs that either in some cases represent a 
high percentage of the restricted samples or that they properties are placed in between 
the values of the extreme pure samples, blurring the differences in the restricted sample.

\section{Summary}
In this work we aimed to analyse whether the differences reported 
in the literature between compact groups identified in different bands 
are still present when a single finding algorithm is applied on samples
of galaxies selected in different photometric bands. 
Therefore, we worked with the synthetic galaxies from the galaxy 
lightcone built by \cite{Henriques_2012}, 
which combines the galaxies from the semi-analytic model of 
galaxy formation of \cite{Guo_2011} and the dark-matter 
particles of the Millennium Run Simulation \citep{springel05}. 
We note that adopting any specific semi-analytic model could introduce a
dependence of the results on the particular set of parameters
and physical processes that were used in the model construction.
Nevertheless, this lightcone is one of the very few freely available mock catalogues 
that provides apparent magnitudes in nine photometric bands obtained for galaxies
extracted from a high resolution N-body simulation, which makes this sample ideal for 
performing comparative studies between properties of compact groups in different bands.
Moreover, this particular semi-analytic model has proven to be quite efficient
at reproducing the observed abundance of low redshift galaxies over a wide range of 
stellar masses and luminosities, and also
the large-scale clustering as a function of stellar masses and galaxy colours \citep{Guo_2011}.
Differences with observations have been indeed reported for the population of galaxies at $z\ge1$, which is 
far outside the redshift range of interest in this work.
Analysing the differences in mock compact groups caused by using different semi-analytic models is beyond 
the scope of this work, and it has been previously assessed by \cite{Diaz_Gimenez+10}.

From the original lightcone, we built three mock catalogues in different
photometric bands (K, r and u) by selecting galaxies brighter than an given 
apparent magnitude limit. We adopted the K-band apparent magnitude limit 
to mimic the observational sample of galaxies of the 2MASS ($K_{\rm lim}= 13.57$). 
By analysing the magnitude-magnitude distribution of galaxies in the lightcone, 
we determined the limits in the other two bands to be equivalent 
to the limit in the K-band ($r_{\rm lim}=16.4$ and $u_{\rm lim}=18.73$). 

We identified compact groups in each of the three mock catalogues by applying 
an automatic Hickson-like algorithm. 
At first, this algorithm identifies compact groups in projection. 
The criteria include membership, compactness, isolation and flux limit. 
The limiting values for each of these criteria have to be changed according to 
the photometric band in order to obtain similar results. 
Once the compact groups have been selected in 
projection, the algorithm performs a velocity filtering to avoid many 
interlopers. We found 447 K-CGs, 406 r-CGs and 276 u-CGs.  

The comparison of properties of CGs identified in the three bands 
revealed that K-CGs and u-CGs present smaller projected radii than r-CGs, 
however, analysing the 3D comoving interparticle separations between the four closest members, 
the u-CGs have the longest separations while the K-CGs have the shortest. 
We do not find differences in crossing times between the three samples of 
CGs. Most of these results are in conflict with previous comparisons reported
in the literature \citep{Diaz_Gimenez+12,herfer15}. The main reason for these
discrepancies is due to the difficult task of comparing groups samples
identified with different algorithms (Hickson-visual inspection or FoF like) 
that could bias the resulting group properties, particularly the sizes which are 
part of the finding criteria.

We also found that the brightest galaxy in u-CGs tends to be brighter than the 
brightest galaxies in r- or K-CGs. The magnitude gap between the two brightest 
galaxies in a group is larger for K-CGs than for u-CGs. The r-CGs present the highest
values of surface brightness in any of the three bands, which means a lowest
``compactness''. 

We have checked if these results were dependent on the different choices of the 
band-dependent limiting values in the identification criteria. 
Therefore, we defined ``restricted'' samples of CGs in the three bands 
by restricting the samples using also the limiting values of the other 
two bands (e.g., the restricted r-CGs accomplish the $\mu_r$ and $r_{\rm bri}$ criteria, 
and also the $\mu_K$, $\mu_u$, $K_{\rm bri}$ and $u_{\rm bri}$ criteria). 
We confirmed that u-CGs are the smallest in
projection, the K-CGs have the largest magnitude gaps and the smallest 3D interparticle 
separations, while the r-CGs have the highest surface brightness and the largest projected sizes. 
Several of these results have been enhanced in the restricted sample in comparison with 
the results obtained for the total samples.

Although we intended to normalise the criteria, the samples of CGs are still different. 
Not all the CGs identified in one band are also present in the other bands. 
In order to disentangle more specifically which differences are intrinsically due to the 
photometric band in which the galaxies were observed, from the samples of restricted CGs we 
selected those groups that are common in the three identification and those that only exist in one 
of the mock catalogues (pure CGs).
The common CGs are half of the restricted u-CGs ($51\%$), while they represent only 1/3 of the K-CGs, 
and we found a percentage in between for the r-CGs ($41\%$). 
Groups that are only identified in the K-band represent 39\% of the restricted K-CGs, 
while the percentage of ``pure-r'' CGs represent $17\%$ of the restricted r-CGs,
and 23\% of the restricted u-CGs are identifiable only in the u-band. 

The comparison between these samples indicates that pure-r CGs are the largest in 
projection, and they also have the highest surface brightness (less compact). 
The pure-u CGs have the brightest first ranked galaxies, 
and the smallest differences between the first and the second ranked galaxies. 
The pure-K CGs have the highest compactness and the smallest virial masses, and mass-to-light ratios.
More noticeable, this sample presents the largest magnitude gaps between the two brightest group
members when compared with all the sample of CGs used in this work. 
This result is a clear indication that this 
characteristic is inherent of groups only identified in this particular photometric band. 
This result is also related with the very low values obtained for the Tremaine-Richstone statistics, 
T1 and T2, which are commonly thought as indication of galaxy mergers.   
Finally, the CGs that are in common in the three bands present the smallest 3D comoving galaxy 
separations between the four closest galaxies, 
which makes them very compact physical entities in 3D space, therefore, 
this sample shows the largest percentage of Real CGs (using \citealt{Diaz_Gimenez+10} definition of Reals). 

Our results indicate that the comparison of compact groups from different sources has
to be done carefully to avoid introducing biases related to the different selection functions, 
but we also demonstrated that there are indeed intrinsic features that differ from band to band. Some of those
differences can be blurred in the bulk of data given the existence of compact groups common 
to all the photometric bands. 
Despite these results were obtained from one single semi-analytic model of galaxy formation,  
this model reproduces very well the observable local luminosity function 
in the $K$, $r$ and $u$ photometric bands, and also the clustering
of galaxies as a function of stellar masses and colours \citep{Guo_2011,Henriques_2012}. Therefore,  
we are quite confident that the differences we found between the mock CGs in different bands
 might also mimic the differences between observational CGs.
However, the predictions presented in this work need to be confirmed from
unbiased studies performed on multi-band observational catalogues, 
and/or when different models of galaxy formation with information 
in multiple photometric bands are released. 
 
\section*{Acknowledgements}
{\small 
The Millennium Simulation databases used in this paper and the web application 
providing online access to them were constructed as part of the activities of 
the German Astrophysical Virtual Observatory (GAVO). 
We thank Bruno Henriques for allowing public access for the outputs of his all-sky mock catalogue 
and kindly answering questions about the sample. 
This work has been partially supported by Consejo Nacional 
de Investigaciones Cient\'\i ficas y T\'ecnicas de la Rep\'ublica Argentina 
(CONICET), PIP: 11220130100365, and the Secretar\'\i a de Ciencia y Tecnolog\'\i a de la 
Universidad de C\'ordoba (SeCyT), 203/14.}

\bibliography{biblio}
\appendix
\section{Tables for pure and common CGs}
\label{apen}
In this section we quote tables including the medians and 95\% confidence intervals for
all the properties under study and the p-values obtained from the comparison among different
photometric bands for the samples of pure and common CGs defined in Sect.~\ref{purmix}.

\begin{table}
\centering
\scriptsize
\caption{Group properties for the samples of pure and common CGs\label{pure_mix} }
\tabcolsep 1.3pt
\begin{tabular}{|c|r|r|r|r|}
\hline
 Sample & $|K|$ & $|r|$ & $|u|$ & $C$ \\
 \# of CGs &         149 &          48 &          52  & 118 \\
\hline
                          $\theta_G$ & $ 2.7 \ (  0.3)$ & $ 4.0 \ (  0.5)$ & $ 4.0 \ (  0.7) $& $ 3.5 \ (  0.3)$  \\
                               $r_p$ & $  63 \ (    6)$ & $  87 \ (   11)$ & $  64 \ (    8) $& $  65 \ (    6)$  \\
                       $R_{\rm vir}$ & $  71 \ (    8)$ & $ 115 \ (   19)$ & $  90 \ (   17) $& $  95 \ (    8)$  \\
             $\langle d_{ij}\rangle$ & $  71 \ (    7)$ & $  93 \ (   13)$ & $  69 \ (   10) $& $  70 \ (    6)$  \\
                           $s_\perp$ & $ 105 \ (   10)$ & $ 125 \ (   19)$ & $ 103 \ (   14) $& $  93 \ (   10)$  \\
                       $s_\parallel$ & $ 123 \ (   23)$ & $ 128 \ (   59)$ & $ 220 \ (  139) $& $ 108 \ (   26)$  \\
                               $s_4$ & $ 173 \ (   24)$ & $ 214 \ (   53)$ & $ 229 \ (  133) $& $ 153 \ (   26)$  \\
                          $\sigma_v$ & $ 255 \ (   27)$ & $ 262 \ (   60)$ & $ 268 \ (   47) $& $ 269 \ (   24)$  \\
                 $H_0 \, t_{\rm cr}$ & $ 2.7 \ (  0.4)$ & $ 2.9 \ (  1.1)$ & $ 2.5 \ (  0.7) $& $ 2.5 \ (  0.4)$  \\
                ${\cal M}_{\rm vir}$ & $ 5.0 \ (  1.1)$ & $ 8.7 \ (  5.8)$ & $ 7.6 \ (  2.4) $& $ 6.1 \ (  1.7)$  \\
                       $K_{\rm bri}$ & $10.0 \ (  0.1)$ & $10.1 \ (  0.2)$ & $ 9.7 \ (  0.3) $& $ 9.8 \ (  0.1)$  \\
                       $r_{\rm bri}$ & $12.8 \ (  0.1)$ & $12.9 \ (  0.2)$ & $12.5 \ (  0.3) $& $12.6 \ (  0.1)$  \\
                       $u_{\rm bri}$ & $15.2 \ (  0.2)$ & $15.2 \ (  0.2)$ & $14.7 \ (  0.3) $& $15.0 \ (  0.1)$  \\
                     $\Delta K_{12}$ & $ 1.5 \ (  0.1)$ & $ 1.3 \ (  0.2)$ & $ 1.1 \ (  0.2) $& $ 1.2 \ (  0.1)$  \\
                     $\Delta r_{12}$ & $ 1.4 \ (  0.1)$ & $ 1.3 \ (  0.2)$ & $ 1.2 \ (  0.1) $& $ 1.2 \ (  0.1)$  \\
                     $\Delta u_{12}$ & $ 1.5 \ (  0.2)$ & $ 1.3 \ (  0.3)$ & $ 1.3 \ (  0.2) $& $ 1.1 \ (  0.1)$  \\
                             $\mu_K$ & $21.6 \ (  0.2)$ & $22.5 \ (  0.2)$ & $22.0 \ (  0.3) $& $21.9 \ (  0.2)$  \\
                             $\mu_r$ & $24.5 \ (  0.2)$ & $25.4 \ (  0.2)$ & $24.8 \ (  0.2) $& $24.7 \ (  0.2)$  \\
                             $\mu_u$ & $26.9 \ (  0.2)$ & $27.6 \ (  0.2)$ & $27.0 \ (  0.2) $& $27.2 \ (  0.2)$  \\
                             $L_K$ & $   231 \ (   23)$ & $ 197 \ (   68)$ & $ 196 \ (   32) $& $ 224 \ (   24)$  \\
                             $L_r$ & $    66 \ (    6)$ & $  63 \ (   18)$ & $  58 \ (    9) $& $  64 \ (    7)$  \\
                             $L_u$ & $    41 \ (    5)$ & $  40 \ (   15)$ & $  41 \ (    7) $& $  41 \ (    4)$  \\
              ${\cal M}_{\rm v}/L_K$ & $  22 \ (    4)$ & $  35 \ (   17)$ & $  34 \ (   10) $& $  27 \ (    5)$  \\
              ${\cal M}_{\rm v}/L_r$ & $  73 \ (   16)$ & $ 126 \ (   55)$ & $ 117 \ (   32) $& $  98 \ (   20)$  \\
              ${\cal M}_{\rm v}/L_u$ & $ 119 \ (   26)$ & $ 183 \ (   93)$ & $ 175 \ (   58) $& $ 143 \ (   34)$  \\
                              $T1_K$ & $0.53  \ (0.04)$ & $0.53  \ (0.07)$ & $0.67  \ (0.09) $& $0.55  \ (0.04)$  \\
                              $T2_K$ & $0.60  \ (0.04)$ & $0.60  \ (0.07)$ & $0.72  \ (0.08) $& $0.58  \ (0.04)$  \\
                              $T1_r$ & $0.50  \ (0.03)$ & $0.56  \ (0.07)$ & $0.62  \ (0.07) $& $0.57  \ (0.05)$  \\
                              $T2_r$ & $0.55  \ (0.03)$ & $0.60  \ (0.06)$ & $0.61  \ (0.07) $& $0.62  \ (0.05)$  \\
                              $T1_u$ & $0.56  \ (0.04)$ & $0.71  \ (0.06)$ & $0.72  \ (0.08) $& $0.58  \ (0.05)$  \\
                              $T2_u$ & $0.62  \ (0.04)$ & $0.73  \ (0.08)$ & $0.59  \ (0.07) $& $0.66  \ (0.05)$  \\
                            \% Reals & $53.7$ & $45.8$ & $40.4 $ & $59.3$ \\
\hline
\end{tabular}
\parbox{8cm}{
\small
{\ct {\bf Notes:} In each cell, the format $xx \ (ss)$ contains the median ($xx$) and the shift ($ss$) to construct an approximated 95\% confidence interval, $CI = xx \pm ss$ (see text for details),
except for the $T_1$ and $T_2$ values where the quantity in parentheses
are the error bars computed using the bootstrap resampling technique. 
Units: $\theta_G$ = $arcmin$; $r_p$, $R_{\rm vir}$, $<d_{ij}>$, $s_\perp$, $s_\parallel$ and $s_4$ = $kpc \ h^{-1}$; $\sigma_v$ = $km \ s^{-1}$ 
; $H_0 \, t_{\rm cr}$ = $10^{-2}$ ; ${\cal M}_{\rm vir}$ = $10^{12} \ M_{\odot} \ h^{-1}$ ;magnitude gaps are calculated in absolute magnitudes 
for each photometric band; $\mu$ = $mag/arcsec^2$; $L$ = $10^{9} \ L_{\odot} \ h^{-2}$; ${\cal M}_{\rm vir}/L$ = $M_{\odot}/L_{\odot} \ h$.}
}
\end{table}
\begin{table}
\centering
\scriptsize
\caption{P-values of the Kolmogorov-Smirnov two sample test. Each column combines different
pairs of pure and common CG samples. Grey shaded cells are included to highlight p-values lower than 0.05.\label{KS_pure_mix} }
\tabcolsep 1.3pt
\begin{tabular}{|c|c|c|c|c|c|c|}
\hline
 & $|K|$ - $|r|$ & $|K|$ - $|u|$ & $|u|$ - $|r|$ & $|K|$ - $C$ & $|r|$ - $C$ & $|u|$ - $C$ \\
\hline
                     $\theta_G$ &$     \ct 2\x 10^{-4}$&$  \ct         0.02  $&$              0.43  $&$    \ct 3\x 10^{- 3}$&$              0.23  $&$              0.21  $\\
                          $r_p$ &$    \ct 2\x 10^{-3 }$&$              0.58  $&$ \ct          0.01  $&$              0.10  $&$    \ct 7 \x 10^{-3}$&$              0.95  $\\
                  $R_{\rm vir}$ &$    \ct 2\x 10^{-4 }$&$              0.12  $&$              0.09  $&$    \ct 3\x 10^{- 5}$&$              0.20  $&$              0.14  $\\
        $\langle d_{ij}\rangle$ &$    \ct 1\x 10^{-3 }$&$              0.35  $&$    \ct 4\x 10^{- 3}$&$              0.17  $&$    \ct 7 \x 10^{-4}$&$              1.00  $\\
$s_\perp$ 			&$    \ct       0.04  $&$              0.62  $&$              0.20  $&$              0.60  $&$    \ct 4 \x 10^{-3}$&$              0.36  $\\
$s_\parallel$ 			&$              0.82  $&$     \ct 4\x 10^{-3}$&$              0.13  $&$              0.88  $&$              0.44  $&$    \ct 5 \x 10^{-3}$\\
                          $s_4$ &$              0.28  $&$    \ct       0.02  $&$              0.13  $&$              0.66  $&$              0.16  $&$\ct           0.01  $\\
                     $\sigma_v$ &$              0.77  $&$              0.73  $&$              0.93  $&$              0.38  $&$              0.28  $&$              0.30  $\\
 $H_0 \, t_{\rm cr}$ 		&$              0.09  $&$              0.83  $&$              0.35  $&$              0.15  $&$              0.34  $&$              0.97  $\\
${\cal M}_{\rm vir}$ 		&$   \ct        0.02  $&$              0.27  $&$              0.26  $&$              0.20  $&$              0.12  $&$              0.84  $\\
                  $K_{\rm bri}$ &$              0.15  $&$              0.13  $&$     \ct      0.02  $&$              0.20  $&$\ct           0.03  $&$              0.39  $\\
                  $r_{\rm bri}$ &$              0.22  $&$              0.08  $&$     \ct      0.03  $&$              0.06  $&$\ct           0.04  $&$              0.39  $\\
                  $u_{\rm bri}$ &$              0.84  $&$   \ct        0.03  $&$     \ct      0.04  $&$              0.28  $&$              0.59  $&$              0.08  $\\
                $\Delta K_{12}$ &$              0.12  $&$              0.07  $&$              0.20  $&$    \ct 3 \x 10^{-3}$&$              0.49  $&$              0.97  $\\
                $\Delta r_{12}$ &$              0.18  $&$  \ct         0.01  $&$              0.66  $&$    \ct 1 \x 10^{-4}$&$              0.22  $&$              0.42  $\\
                $\Delta u_{12}$ &$              0.18  $&$  \ct         0.01  $&$              0.66  $&$    \ct 1 \x 10^{-4}$&$              0.22  $&$              0.42  $\\
                        $\mu_K$ &$    \ct 2 \x 10^{-5}$&$   \ct        0.03  $&$    \ct 7\x 10^{- 3}$&$              0.07  $&$    \ct 2 \x 10^{-4}$&$              0.34  $\\
                        $\mu_r$ &$    \ct 6 \x 10^{-5}$&$              0.06  $&$    \ct       0.01  $&$              0.15  $&$    \ct 2 \x 10^{-4}$&$              0.34  $\\
                        $\mu_u$ &$    \ct 5 \x 10^{-4}$&$              0.22  $&$    \ct 9\x 10^{- 3}$&$              0.10  $&$    \ct 2 \x 10^{-3}$&$              0.79  $\\
          $L_K $ 		&$              0.39  $&$              0.39  $&$              0.83  $&$              0.69  $&$              0.11  $&$              0.42  $\\
          $L_r $ 		&$              0.53  $&$              0.56  $&$              0.68  $&$              0.73  $&$              0.26  $&$              0.31  $\\
          $L_u $ 		&$              0.30  $&$              0.80  $&$              0.46  $&$              0.66  $&$              0.08  $&$              0.60  $\\
         ${\cal M}_{\rm v}/L_K$ &$    \ct       0.03  $&$              0.07  $&$              0.67  $&$              0.29  $&$              0.06  $&$              0.44  $\\
         ${\cal M}_{\rm v}/L_r$ &$    \ct       0.04  $&$              0.11  $&$              0.67  $&$              0.29  $&$              0.21  $&$              0.59  $\\
         ${\cal M}_{\rm v}/L_u$ &$              0.06  $&$              0.20  $&$              0.64  $&$              0.20  $&$              0.26  $&$              0.96  $\\

\hline
\end{tabular}
\end{table}

\label{lastpage}
\end{document}